# The QUEST Large Area CCD Camera

C. Baltay, D. Rabinowitz, P. Andrews, A. Bauer, N. Ellman, W. Emmet, R. Hudson,
T. Hurteau, J. Jerke, R. Lauer, J. Silge, A. Szymkowiak

Yale University

B. Adams, M. Gebhard, J. Musser

Indiana University

M. Doyle, H. Petrie, R. Smith, R. Thicksten

California Institute of Technology

J. Geary

Harvard Smithsonian Center for Astrophysics

## ABSTRACT

We have designed, constructed and put into operation a very large area CCD camera that covers the field of view of the 1.2 m Samuel Oschin Schmidt Telescope at the Palomar Observatory. The camera consists of 112 CCDs arranged in a mosaic of four rows with 28 CCDs each. The CCDs are 600 x 2400 pixel Sarnoff thinned, back illuminated devices with 13_ x 13_ pixels. The camera covers an area of $4.6^o$ x $3.6^o$ on the sky with an active area of 9.6 square degrees. This camera has been installed at the prime focus of the telescope, commissioned, and scientific quality observations on the Palomar-QUEST Variability Sky Survey were started in September of 2003. The design considerations, construction features, and performance parameters of this camera are described in this paper.

**Keywords:**

Instrumentation: Detectors
Techniques: Photometric
Survey
Astrometry

# **CONTENTS**

## 1. INTRODUCTION

Schmidt telescopes are the instrument of choice for large area sky surveys because of their large field of view. However, the telescopes have very large, curved image planes and are difficult to fully instrument with silicon detectors. Until recently these telescopes have been used for large area sky surveys with photographic plates (Reid et al. 2002). A few years ago the Near Earth Asteroid Team (NEAT) instrumented the Palomar Samuel Oschin Schmidt Telescope with three 4080 x 4080 CCDs (Pravdo et al. 2001). This camera covered an area of 3.75 square degrees. Recently our QUEST collaboration has designed, built, and operated a large area camera (Baltay et al. 2002) consisting of 16 CCDs 2048 x 2048 pixels each on the 1 m Schmidt Telescope at the Venezuelan National Astronomical Observatory (CIDA). This camera, covering 5.4 square degrees, was the prototype of the camera described in this paper.

The Very Large Area QUEST Camera was designed to operate at the 48" Samuel Oschin Schmidt Telescope (Harrington 1952) at the Palomar Observatory. The properties of this telescope are summarized in Table 1. The camera is located at the prime focus inside the telescope tube as shown in Figure 1.

The camera consists of 112 CCDs arranged in four rows or "fingers" of 28 CCDs each as shown in Figure 2, and covers 4.6° by 3.6° (north-south by east-west) on the sky. We label the fingers A,B,C, and D and the columns of CCD's 1-28 as shown.

The gaps between the active areas of the CCDs in a row are typically 1 mm, and between the active areas in adjacent rows the gaps are typically 22.8 mm. The exact location of each CCD in the array varies slightly with declination as the four rows of CCDs are rotated with respect to each other, as described below. The effective area covered by the active areas of the CCDs in the camera is 9.6 square degrees. The properties of the camera are summarized in Table 2, and a photograph of the camera is shown in Figure 3.

The camera has been designed to operate in the drift scan mode (McGraw et al. 1980), which is also referred to as time-delay integration mode (Gibson and Hickson 1992). In this mode, the telescope is locked into a fixed position at a given declination angle. The CCD array is oriented with the columns of pixels in the clocking direction lined up precisely in the east-west direction, and the CCDs are clocked synchronously with the motion of the star images across each CCD. Each star image thus crosses four CCDs, one in each row of CCDs. Each row of CCDs can have a filter of a different color in front of it so that the camera can collect images in each of four colors essentially simultaneously. The camera thus has a 100% duty cycle (i.e., data are collected continuously), and no time is lost as a result of the readout time or telescope slewing time. Since the telescope is locked into a fixed position and is not tracking, the system is very stable and produces more accurate photometric measurements. Photometric precision is further enhanced because each point in the sky is imaged by averaging over an entire column of pixels, and thus pixel-to-pixel variations in sensitivity are minimized.  In this mode during a clear night a strip of sky 4.6° in declination by approximately 120° in right ascension, or about 500 square degrees, can be covered in each of four colors.  The effective exposure time on each CCD is governed by the rotation rate of the Earth and is typically 140 seconds, depending slightly on the declination.

The camera and its control system are also able to operate in a conventional point and track mode which is useful in many applications where a different exposure time is desirable.

The scientific motivation for building this camera was to carry out a variability sky survey of a large area of the sky.  In a five year period we plan to cover an equatorial strip between +25° and -25° in declination at all right ascensions, an area of ~ 15,000 square degrees outside of the galaxy, repeatedly covering multiple times each at time intervals varying between minutes to years with multiple color bands.  This variability survey is being utilized by a variety of science projects ranging from high redshift quasar surveys, gravitational lensing of quasars, supernova studies, to searching for Kuiper belt objects and near earth asteroid tracking.

In the following sections we will describe the principle of operation of the camera (§2), the construction of the camera hardware (§3), the readout electronics (§4), the robotic control and operation of the instrument (§5), the data transfer, processing, and archiving (§6), and the performance of the instrument (§7).

## 2. PRINCIPLES OF OPERATION OF THE CAMERA

If the camera were used to drift-scan along the equator, the images of stars would follow straight lines and move at the same rate across the image plane. However, at declinations other than the equator, the stars will very nearly follow arcs of circles, and stars at different north-south positions will move at different rates. In drift-scanning, the sagittae due to the first effect will smear the images in the north-south direction, and due to the second effect, some of the stars will not be exactly synchronous with the CCD clocking rates and thus will be smeared in the east-west direction. In order to keep these effects at an acceptable level, i.e., to keep the smearing of the point-spread function (PSF) below about 1" in any direction, we rotate each CCD by an amount dependent on the declination being scanned in such a way that the clocking direction of each CCD is tangential to the arcs that the stars are moving in at that location in the array. This is accomplished by mounting each of the four rows of 28 CCDs in a north-south row on an Invar bar, which we call a "finger." Each of the four fingers can be rotated by a different amount by cams that are driven by external, computer-controlled stepper motors (see Figure 4). An exaggerated sketch of this scheme is shown in Figure 5.  In addition, each column of CCDs is scanned along a slightly different declination, and therefore the parallel clocks reading out the CCDs are synchronized at slightly different rates.

The radius $r_\delta$ of the star tracks (i.e., the arcs of circles along which the image of a star moves in drift-scanning, Figure 5) on the image plane of a telescope with focal length f at a declination $\delta$ is to a good approximation given by

$$r_\delta = \frac{f}{\tan \delta}.$$

The parallel clock rate __ for reading out a CCD in such a way that the motion of the charge is synchronous with the motion of the star image across the CCD at a declination $\delta$ is given by

$$v_\delta = \frac{\Omega f}{a} \cos \delta,$$

where $\Omega = 72.7$ μrad s$^{-1}$ is the rotation rate of the Earth, f is the focal length of the telescope, and $a$ is the pixel size on the CCD. For the Palomar Schmidt telescope and our pixel size, this gives

$$v_\delta = 17.1 \cos \delta \; lines \; s^{-1}.$$

In drift-scanning along the equator, the readout parallel clocks are thus synchronized at approximately 17 lines s$^{-1}$. At this rate, a star image takes

$$t = \frac{140.2}{\cos \delta} \text{ seconds}$$

to cross a CCD. This is the effective integration or exposure time in drift scan mode. This exposure time is governed by the rotation of the Earth and cannot be changed. However, since each star crosses four CCDs, these can be added for an effective exposure time of about 560 s. In cases where even longer exposure times are desirable, repeated scans of the same area of sky can be performed and co-added.

The angle by which the CCD support fingers (see Figure 5) have to be rotated to keep the clocking direction of the CCDs tangent to the star tracks on the image plane at a declination $\delta$ is

$$\Delta \theta = \frac{d}{f} \tan \delta$$

where d is the distance of the pivot point of each finger from the camera center line ($d = \pm 8.25$ cm for fingers 1 and 4 and $\pm 2.75$ cm for fingers 2 and 3) and $\delta$ is the declination at the pivot point. Thus, for example, the top finger (1) has to be rotated about $0°.15$ for $\delta = 6°$, which is very small but nevertheless quite important to keep the image sizes small. As mentioned above, the columns of CCDs scan along slightly different declinations and thus have to be clocked at slightly different rates. The detailed implementation of this scheme is to clock all four columns at the same rate but drop clock pulses at different rates in different columns to produce the average clock rates required. The fast serial readout clock is 100 kHz for all of the CCDs in the array.

The scheme described above for varying the rotation and clock rate synchronization of the different CCDs in the array removes the dominant effects that smear the images. There are, however, residual effects due to the sagitta of the image motion and spread in the rate of motion of the images across the finite width of a single CCD, as illustrated in Figure 6. For a CCD with length $l$ (in the east-west direction) and width w (in the north-south direction), the residual smearing of the image sizes $\Delta x$ and $\Delta y$, in the north-south and east-west directions, respectively, scanning at a declination $\delta$, is given by

$$\Delta x = \frac{1}{8} \frac{l^2}{f} \tan \delta,$$

$$\Delta y = \frac{1}{2} \frac{lw}{f} \tan \delta,$$

where f is the focal length of the telescope. This residual smearing limits the range of declinations at which this camera can be used in the drift-scanning mode without intolerable degradation of the image sizes. It is this consideration that led to the narrow (600 pixel wide) format for the CCDs.

The design has been optimized in such a way that the residual image smearing is kept below 1" for declinations up to $\pm 25°$. Given the typical seeing at Palomar site, we can drift-scan at declinations up to $\pm 25°$ with no appreciable degradation of image quality. This is sufficient for the equatorial survey for which the camera has been designed. Of course, the camera can be operated in a conventional point-and-track mode to cover regions of the sky above these declinations.

Another complication of the design is due to the fact that the image plane of a Schmidt telescope is not flat but a spherical surface concentric with the primary mirror. To arrange the CCDs in such a shape would have been cumbersome. Instead, we designed and had built a 36 cm diameter field flattener lens that covers the entire image plane. This lens produces a flat image plane and, in addition, corrects for the pincushion distortion inherent in the telescope to a level where the degradation of the image shapes is negligible.

The depth of field of the Palomar Schmidt telescope is quite shallow. For this reason, the front surfaces of the 112 CCDs, including the motion of the finger mounts, has to be kept in a plane to a tolerance of less than $\pm 25 \mu m$. It required great care in the machining and the alignment procedures to achieve this precision.

During the commissioning period, after the camera had been installed in the telescope, a great deal of effort was expended to align the plane of the CCDs with the focal plane of the telescope. Once this had been achieved, however, it was quite stable and required no further adjustment. Typically, before each night's data-taking, the focus of the telescope, the rotational position of the fingers, and the synchronized readout rate, which have been set by the control computers for the appropriate declination, are checked by looking at the shape and size of stellar images.

**3. DESCRIPTION OF THE QUEST CAMERA**

The CCD camera is located at the prime focus of the Palomar Schmidt telescope about 3 m from the primary mirror, with the CCDs facing the mirror. The outside dimensions of the camera body had to be kept small to fit mechanically at the prime focus and to obscure as little of the incoming light beam as possible. The CCDs inside the camera have to be cooled to $-100^{o}C$ to minimize the dark current. The outside of the camera has to be kept at ambient room temperature to prevent thermal gradients and turbulence inside the telescope. These requirements complicated the design of the camera, requiring a vacuum system for thermal insulation, a cryogenic cooling system using liquid nitrogen, a nitrogen filling system, electrical feed throughs through the vacuum dewar, etc. These systems will be briefly described in this section.

3.1. The CCDs

The heart of the camera is the mosaic of 112 CCDs arranged in four rows as shown in Figure 2. Each CCD has 600 x 2400 pixels, 13 _m x 13 _m each, as shown in Figure 7. The reason of the narrowness of the CCDs (600 pixels in the north-south direction) is to keep the image smearing due to the need to vary the clocking rate with declination at an acceptable level, as already discussed in Section 2 above. The CCDs were thinned to about 15 _m thickness and used in a back side illuminated mode. This feature was driven by the desire to achieve a high quantum efficiency across a wide range of wavelengths, and to have some appreciable quantum efficiency in the blue and ultraviolet bands.

There were no commercially available CCDs that satisfied our requirements so we had to design our own CCDs for this camera. The CCDs were fabricated to our design by Sarnoff Corporation's silicon fab house. The production process consisted of two fab runs with 24 four inch wafers each, with 12 CCD devices on each wafer. The packaging for the CCDs was designed and carried out at Yale. A schematic drawing of the packaged CCD is shown in Figure 8. Of the total of the 576 CCDs, 385 were packaged and tested at Yale, and the best 112 were selected and installed in the camera.

The thinned (15 _m thick) CCDs were glued to a 500 _m thick glass plate (Hoya NA-40 glass with good UV transmission) at the wafer level and the devices were diced at Sarnoff together with the glass. All of the 18 contact pads on each CCD are along one narrow edge; these were wire bonded directly to pads on the flexible cable which leads to the connectors on the far end. The rest of the package was then assembled as shown in Figure 8.

To further enhance the quantum efficiency an antireflection coating (450 nm of bismuth oxide) was applied to the illuminated surface of the CCDs. The quantum efficiency of the packaged devices was measured at Indiana University for part of the wavelength range. These measurements, compared with the design goal, are shown on Figure 9. The dark current of the CCDs varied from device to device. The dependence of the dark current on the temperature for two typical devices is shown in Figure 10.

The read noise of the CCDs have been measured to be 6.5 electrons/readout. The distribution in the read noise for a typical CCD is shown in Figure 11.
The properties of these CCDs are summarized in Table 3.

3.2. The Field-Flattener Lens

The Palomar Schmidt normally has a focal plane that has the shape of a convex spherical surface. To allow the CCD array to be in a flat plane, we have designed a field-flattener lens to reimage to a flat focal plane. Great care had to be exercised in this design to keep pincushion and other distortions below a few microns in the image plane to allow the camera to be used in a drift-scan mode. The lens was designed by the QUEST collaboration and was manufactured by Coastal Optical Systems of Riviera Beach, Florida. The lens has a 36 cm diameter and is also used as the vacuum window at the front of the detector. The lens is biconvex, with two spherical surfaces of radius of curvature of 3635 and -1367 mm, respectively. It is 25 mm thick in the center. Careful finite-element stress analysis calculations had been carried out to show that the lens is strong enough to serve as a vacuum window and the deflection is small enough not to distort the optical properties of the lens unduly. The lens is made of fused quartz, Corning 7980, which has good transmission in the ultraviolet as well as over the whole required wavelength range. The lens does not have an antireflection coating. The lens is located about 2 cm in front of the CCD image plane. We have examined the image sizes and shapes at different locations on the image plane. No variation or degradation as a function of distance from the center of the image plane has been observed. We therefore believe that this lens has flattened and corrected the field to an acceptable level.

3.3. The Camera Dewar

To be able to cool the CCDs to -100° C, they have to be in a vacuum enclosure. The detector housing is thus a vacuum vessel cylindrical in shape, 18 inches in diameter, and about 5 inches deep as shown in Figure 12. The front face serves as the field-flattener lens discussed above. The back plate of the housing has the mounting bracket on it to attach the detector to the telescope. The back plate also has penetrations for the vacuum pump port, cold fingers to cool the CCDs (see section 3.4), and all of the electrical feedthroughs. The housing is made of aluminum, polished on the inside to reduce radiative heat loss, and anodized flat black on the outside to reduce reflections and glare.

### 3.4. The Cooling System

To keep the dark current at an acceptable level the CCDs are typically cooled to $-100^{\circ}$C. The cooling is done by liquid nitrogen. The heat load on the camera, dominated by the heat radiated in through the large vacuum window and generated by electronics inside the dewar, is about 50 watts. This requires typically 2 liters of liquid nitrogen per hour. In most conventional liquid nitrogen cooled cameras the camera contains a dewar for the liquid nitrogen which is filled typically daily. In our case this solution was not possible. The shallow space allowed for the camera between the focus hub and the focal plane did not allow enough room for the large dewar required. We therefore placed a 10 liter dewar on the other side of the focus hub, with two narrow vacuum insulated tubes extending to the back of the camera. The liquid nitrogen flows to the end of these tubes where they meet copper cold fingers which extend inside the camera and are attached in a flexible way to the fingers that the CCDs are mounted on. The dewar is kept half filled with liquid nitrogen by an automatic filling system from a large liquid nitrogen dewar located outside of the dome. This arrangement is shown schematically in Figure 13.

Heaters and temperature sensors are glued to the backside of the invar fingers upon which the CCDs are mounted. Feedback circuits control the current in the heaters to keep the CCDs at $\sim -100^{\circ}$C.

### 3.5. The Vacuum System

The camera dewar is operated at a vacuum of $\sim 10^{-4}$ torr, which is sufficient to keep the convective heat loss at a negligible level. The vessel is evacuated by a turbomolecular pump preceded by a roughing pump. These pumps are not connected during observations. In fact, once the vessel is evacuated, the vacuum usually lasts for several months, so the pumps are connected only periodically (typically once every few months).

### 3.6. The Focal Plane and CCD Supports

The CCD packages are mounted on four Invar bars (fingers) each 0.25 inches thick, 1.0 inches wide, and 12 inches long. These four fingers in turn are attached to a 16 inch diameter Invar base plate 0.5 inches thick (see Figure 4). This plate has a large rectangular opening in the center to allow the electrical connections from the CCDs to pass through to the preamplifiers and clocking electronics located at the back of the dewar under this plate. To keep the focal surface flat to the required 25 μm tolerance, all of the Invar bars and the support plate were ground to 5 _m precision, and all joints are spring-loaded. The Invar parts were coated with an Armoloy coating to prevent corrosion. After the entire system was assembled, a detailed optical survey showed that the front surfaces of all of the CCDs were in a plane with an rms scatter of 11 _m.

The Invar fingers pivot at one end, and their rotational position is controlled by cams located near the other end under the Invar support plate. The shafts of the four cams penetrate the back plate of the dewar to the gears and stepping motors located outside of the dewar. The Invar support plate is supported from the dewar back plate by three 10 cm long hollow stainless steel tubes to reduce the heat conduction to an acceptable level.

### 3.7. Color Filters

The filter box is located a few centimeters in front of the vacuum window/field-flattener lens. Filter trays can be easily inserted or removed manually. A filter tray consists of four different filters each 5.5 cm wide and 28.0 cm long. Each filter is located in front of one finger of CCDs so that in the course of a drift scan, star images pass through each of the four filters in turn so that data can be collected in four colors simultaneously. A sketch of a filter tray is shown in Figure 14. Several filter trays exist, and the individual filters can be shuffled to make up filter trays in any desired combination. The available filter colors and their wavelength bands are shown in Figure 15. These filters were designed specifically for the QUEST camera, but they resemble the Johnson color system (Bessel 1990) and the Gunn color system (Gunn et al. 1998) quite closely. The optical thickness of the different filters were designed such that the entire system is parfocal. To prevent vapor condensing between the filters and dewar window, dry air is continuously pumped between the surfaces.

### 3.8. The Shutter Box

The camera shutter is located in front of the filter box. The requirements on this shutter are that there should be a 14 inch diameter clear opening but that no part of the shutter box should extend beyond the 18 inch diameter of the camera, since the camera is located at the prime focus of the telescope and obscuration of the light path must be minimized. No such shutter was commercially available, so a shutter was custom-designed and built for this purpose. The shutter consists of two flaps (see Figure 14) which are controlled by a small pneumatic cylinder. The shutter is computer controlled and opens or shuts in about half a second.

## 4. THE CAMERA READOUT ELECTRONICS

The readout electronics for the camera are based on a highly customized version of the MFRONT hardware developed for the Subaru camera (Miyazaki et al. 2002). The specifications of the readout electronics are summarized in Table 4. The 56 front end electronics (CLKAMP) cards are mounted below the focal plane inside the dewar body, and are connected to the CCDs through kapton flex circuits. Each front end card provides video preamplification, clocking switches, and pass through of the bias voltages for two CCDs.  In order to support the line dropping technique used in drift scan mode the vertical clocks from a given front end card are common to two CCDs at the same position on adjacent fingers, while the horizontal clocks are common to two adjacent CCDs on the same finger.

The video outputs of pairs of CLKAMP cards, which contain the circuits to clock the CCDs and the front end amplifiers for the output signals, are multiplexed to the outside world through a 2-1 analog switch.  The digital clock signals for the mosaic are fed through the back wall of the dewar through a custom high density vacuum feed-through.  A second vacuum feed-through is used to pass the video and bias voltages through the dewar wall.

All signals are run to a 6U VME crate mounted to the side of the telescope, which houses controller and digit clock driver cards, and 28 Bias/ADC cards.  Each Bias/ADC card contains two 16 bit ADCs operating at 100 kbs, and a set of programmable DACs that are used to generate the CCD clock rails and dc bias voltages.   Each Bias/ADC card provides 4 CCDs with clock rails and bias voltages.  Each ADC is connected to the output of a 2-1 video switch, enabling each Bias/ADC card to read out 4 CCDs.

A full horizontal read cycle commences with the readout of one row of _ the CCDs in the mosaic (those connected to the first input of the 2-1 video switches), followed by the readout of one row of the remaining CCDs.  A full image can be acquired in point and stare mode through a sequence of horizontal read cycles in roughly 30 seconds.

In drift scan mode, the frequency of the line start clock that initiates a parallel transfer is set to the value determined by the declination.   Differential scan rates in the declination direction are compensated for by dropping parallel transfers (line dropping) at progressively higher rates for CCDs at higher declinations.  Digital clock generation and timing are provided by a controller card based on a Xilinx FPGA which is remotely programmable to allow operating modes and clock timing to be changed under operator control.  The timing of the line drops is controlled by an external, single-board computer.  The controller card packs the digitized data from each transfer into a data frame that is transmitted to an external computer in the control room through an EDT-RCI Gbit fiber link.  All data from the camera are transmitted through a single optical fiber, which runs to the control room.  A separate optically-isolated serial link provides slow control of the system, including voltage setting and monitoring.

A schematic layout of the readout electronics is shown in Figure 16.

## 5. ROBOTIC TELESCOPE AND CAMERA CONTROL AND OPERATION

5.1 Control Programs

To automatically control the telescope and camera, we use many different programs running simultaneously on a network of computers in the control room of the dome (see Figure 17). Each of these programs manages a different piece of hardware. The telescope manager (running on "Pointy") controls the telescope, including pointing, tracking, dome rotation, dome shutter, and focusing. The camera manager (running on "Quest7") controls camera, camera shutter, and drop counter (see Sec 5.3). There is also a manager to control the orientation of the fingers inside the camera and another to control the temperature of the CCDs. All of the programs run under UNIX-based operating systems and receive commands through the UNIX socket protocol (Bach 1986).

There are two additional programs (observation managers) that we use to choreograph the observing operations. One manager (running on Quest16) is for sequencing driftscans and the other (running on Asok) is for point and shoot sequences (Pravdo et al. 2000, 2001). In either case, the manager starts by reading in a script that details the sequence of telescope pointings and camera exposures planned for the night. The program then takes control of the telescope and camera, sending commands to the various hardware managers as necessary to complete the scripted observations. All of the operations of the observation manager can be run interactively at a computer terminal. This is useful for testing the code and for unscripted, manual operation of the telescope and camera.

During the night, the observation manager controls all operations autonomously except for opening the dome. This is the only operation that requires human intervention. After sunset, the operator of the 200" Hale Telescope (where there is always an onsite operator) monitors the weather conditions. If the conditions are good for observing, the operator opens the dome remotely by sending the appropriate command from a computer terminal at the 200" dome. As soon as the observation manager senses that the dome has opened (by queries to the telescope manager), it starts pointing the telescope and taking exposures as scripted. The controller for the telescope is connected to a weather station and will automatically close the dome whenever conditions are bad (rain, high winds, high humidity). However, the 200" operator remotely monitors the dome status and will command the dome to close in bad weather if it does not do so automatically. Once weather conditions improve, the 200" operator must reopen the dome to permit further operations. Until then, the observation manager remains idle. When the scripted observations are complete or when the night ends, the observation manager closes the dome and puts the telescope in stow position for the day.

5.2 Camera Readout

To read out the camera and write the data to disk efficiently we use two computers (see Figure 17). One computer (Quest7) receives digitized image data directly from the camera hardware through a high speed optical-fiber interface. The line transfer rate is accurately controlled by a timer that sends pulses at a fixed rate to the camera. Each pulse initiates the parallel transfer and readout of a single line of pixels from each CCD. Each line is 640 pixels, of which 40 pixels are overscan pixels. Since the CCDs are clocked simultaneously at an average rate of 50 KHz, an entire line from all 112 CCDs consists of 71680 pixels transferred to Quest7 at 5.6 MHz. Within each line, the CCD pixels are interleaved, with one pixel from each CCD, followed by the next pixel from each CCD, and so on.

Nearly as quickly as the data are received by Quest7, they are transferred through a dedicated gigabit network connection to a second computer (Quest16). There the data are handled by two separate programs running simultaneously. One program receives the data from Quest7 over the network, separates the interleaved pixels, and stores the data for each CCD into a separate memory cache. The second program reads each memory cache as it is filled by the first program, and writes the respective contents to a separate file on disk. The first program is able to read the entire CCD-mosaic image (172 million pixels including over scan) into memory on Quest16 in 33 seconds. However, there is an additional overhead of 40 seconds (limited by the speed of the disk controller and network) before the second program completely writes the data to disk. For exposure times less than 40 seconds, the transfer time limits the duty cycle. The total cycle time is a factor of 2 faster than required for drift scanning (140 seconds, see Section 6).

5.3 Parallel Transfer Timing

As discussed in Section 2, drift scanning across a very large field requires that we individually vary the parallel transfer rates of the CCDs as a function of the declination range they cover on the sky. We accomplish this by setting the pulse rate of the timer that controls the parallel transfers (see above) to the highest rate required, $r_o$. This corresponds to the rate required by the row of CCDs closest to the celestial equator. For each CCD row that requires a slower readout rate, $r$, we drop a fraction of these pulses so that the average parallel transfer rate matches the required rate. The integer number of parallel transfers that pass between dropped transfers is the drop count, $n = r/(r_o-r)$. We use a small single-board computer attached to the camera electronics to count the timer pulses and keep track of 28 different drop counts. Whenever a pulse needs to be blocked for a particular column, the computer changes the state of one of 28 respective logic gates within the camera electronics.

5.4 Finger Control

The Finger Controller is a Motion Group MMC-4 microprocessor designed for multiplexing the drive currents for 4 stepping motors. Although the controller is programmable, we use it only to step the motors forward and backward by integral counts, and to put the stepping motors in their home position. Once homed, the finger manager keeps track of the angular position of each finger by summing the total steps forward and backward on each motor. Limit switches on each of the cam axes prevent the motors from exceeding the range of motion allowed by the cams. The home position for each motor is accurately determined by an optical switch that is shut off by a slotted disk on the motor axis. Since each motor turns 8 times on its axis for each rotation of the cam axis, this switch must be logically anded with a contact switch at the home position on each cam axis. We find that there is negligible error in the absolute positioning of motors after repeated cycling to the home position. Even after power cycling the finger controller, absolute positioning is maintained because the finger manager keeps a record of the last position of each motor.

5.5 Temperature Control

The temperature controller consist of 4 i-series temperature-control modules from Omega Engineering, one for each finger in the camera dewar. All four controllers are programmed by the temperature manager through a single serial connection. Each controller senses the temperature on the back side of a finger through a resistive temperature detector (RTD) glued to the back of the finger. To keep the finger temperature constant, the controller varies a 0-10 V control signal that is supplied to a voltage regulator. In response to the control signal, the voltage regulator holds 0-16 V across two 50-ohm power resistors wired in parallel and glued to the back side of the finger on opposite ends. Each resister can thus deliver up to 5 watts of heat to the finger. The controllers use a proportional-integral-differential algorithm (Barr 2002) so that their output signals are optimally varied to keep the finger temperatures constant. Once programmed, the controller modules are autonomous. The function of the temperature manager is only to monitor any fluctuations in the set temperatures.

## 6. DATA RATES, TRANSFER, PROCESSING AND ARCHIVING

6.1. Data Rates, Transfer, and Archiving

The Analog to Digital Converter (ADC) outputs 16 bits or 2 bytes for the signal in ADU (ADC units) from each pixel. The calibration of the electronics varies slightly from CCD to CCD but is in the vicinity of 4 electron/ADU. In the drift scan mode the entire array of $161 \times 10^6$ pixels is read out every 140 seconds for an average data rate of 2.3 Mbytes/sec. The readout control computers store this data on a local disk; during an eight hour night we accumulate 60 Gigabytes of raw data in the drift scan mode, and 100 gigabytes for the point and track mode with 60 second exposures. We currently transfer all data in near real time (within a few minutes of acquisition) to computers at both Yale and Caltech using a radio-frequency link to the internet (HPWREN, Hansen et al, 2002). At Yale we repackage and transmit the data to the Lawrence Berkeley Laboratory (LBL) and the National Center for Supercomputer Applications (NCSA) in Illinois for processing and archival storage.

The raw data is archived on spinning disk storage at Yale and Caltech (we have 15 Terabytes each at this time) and on magnetic tapes at NCSA and LBL. This redundancy provides backup in case of storage equipment failure at any of the institutions.

The software to process such a large volume of data is not trivial. The collaboration has developed several different software pipelines to analyze the data, each optimized for a different science topic: a) the Yale Photometric Pipeline to analyze the data from the drift scan on a daily basis, b) the Deep Search Pipeline, coadding data at the pixel level from all available nights to search for very faint objects, c) the LBL Supernova Pipeline, a highly specialized program to search for supernovae and other variable objects, d) the NEAT Pipeline to search for near Earth Asteroids and Trans-Neptunian Objects, e) the Caltech Data Cleaning Pipeline to remove instrumental artifacts, and f) the Caltech Real Time Pipeline to analyze the data in real time to detect rapidly varying transients. The outputs of these pipelines are typically object catalogs that form the starting points for more specialized data analysis programs. These output object catalogs are also archived at Yale, NCSA, Caltech, and Berkeley.

As discussed above, both the raw data and the output object catalogs are archived on spinning disk and magnetic tapes at several of the collaborating institutions. At this time these archives are in the formats developed for the purposes of our collaboration. We do however have plans to make these archives consistent with the standards of the National Virtual Observatory (NVO, Szalay et al 2001). A schematic of the Data Distribution, Processing and Archiving is shown in Figure 18.

6.2. The Yale Photometry Pipeline

This program was developed specifically for the QUEST drift scan survey (Andrews 2003) with four color filters, currently either Johnson UBRI or Gunn rizz, (we doubled up on the z filter to increase our sensitivity to high redshift objects). This pipeline does some preprocessing (flat fielding, bias subtraction, etc.), astrometry, and both aperture and PSF photometry. The astrometric calibration uses the USNO A2.0 Catalog (Monet 1998) and is typically accurate to 0.1 arcsec. The photometric calibration is done using Landolt (Landolt, 1992) and Stetson (Stetson 2000) Standards as well as the Sloan Digital Sky Survey (SDSS) Data Release 5 Catalog (Adelman-McCarthy et al. 2006). The photometric calibration of 112 CCDs with multiple color filters and varying observing conditions is not easy – at the present it is good to about 5%. More accurate calibrations will be done for specific data analyses as required. The program is run automatically using the computers at NCSA to process the previous night's data. At the present it takes about eight hours on 28 computer nodes to process one entire night of data. This program has been ready for use at the beginning of scientific data with the QUEST camera in September 2003 and we have been successful in keeping the processing up to date.

The software pipelines and the calibration procedures will be described in more detail in a separate article now in preparation.

6.3. The Deep Search Pipeline

The purpose of this program is to search for very faint objects, pushing the depth sensitivity of the data set to its limit (Djorgovski 2003). The program starts with a pixel by pixel coaddition of all of the accumulated observations of any given area of the sky with any particular color filter. Object detection is then carried out on the coadded data, and photometry is performed for all detected objects. With an eventual 10 to 20 scans of each area of the sky with each filter this should represent a considerable improvement in the limiting magnitude we can reach. The program is currently under development and will be described in a later publication.

6.4. The Supernova Pipeline

This program was developed by the Supernova Cosmology Project (Perlmutter et al. 1999) and was adapted to work with the output of the QUEST camera in both the drift scan and the point and track mode. In this program, data from "discovery" nights and from "reference" nights (typically about 2 weeks earlier than the discovery nights) are convolved to the same seeing and normalized to the same intensity scale. The reference nights are then subtracted pixel by pixel from the discovery nights. The vast majority of the objects disappear in the subtraction. Objects with significant residuals in the subtraction are examined visually as candidates for variable objects such as supernovae, Trans-Neptunian objects, asteroids, etc.

6.5. The Real Time Pipeline is now under development at Caltech to process data essentially in real time to detect rapidly varying or transient objects (Mahabal 2003).

7. **PERFORMANCE OF THE INSTRUMENT**

The QUEST camera was installed in the Palomar Samuel Oschin Schmidt Telescope and commissioned in 2003. Routine observations were started in September of 2003. This camera is the only instrument on the telescope and is in use essentially every night. About half of the observing time is devoted to drift scan observations and half to point and track. In a typical lunation 11 to 12 nights are scheduled for each of the two types of observations, with five nights around full moon scheduled for routine maintenance of the telescope or the instrument, or more often devoted to specialized observations that can tolerate the full moon. We find that, averaged over a year, weather conditions are sufficiently favorable to allow data taking about two thirds of the scheduled nights. Of the nights when data are taken about 80% are usable for photometry. The observing time in the drift scan mode is evenly divided between the use of the Johnson UBRI filters and the Gunn rizz filters. The point and track survey uses a broadband red filter RG-610. In the period from September 2003 to July 2005 the Drift Scan Survey covered a total of ~ 15,000 square degrees in the equatorial band between declinations -25° and +25°, covering all right ascensions except for small gaps at the Milky Way. All of this area has been scanned at least four times, twice with the UBRI and twice with the rizz filters. The point and track survey covered 20,000 square degrees repeatedly.

The camera had some problems that had to be accommodated in the software and calibration procedures. A mosaic view of the 112 CCDs of the camera in the drift scan mode is shown in Figure 19.

a) Six of the CCDs do not read out. We believe that this is due to problems in the cables and connectors inside the camera. We decided to live with this for the time being. These problems will be repaired in the next maintenance downtime.

b) Some CCDs had faint light emitting regions near the readout amplifier in one corner. These do not affect the drift scan mode significantly, but these corner areas are not used in the point and track mode.

c) The CCDs have non-linearity response at low illumination (less than 30 or 40 electrons/pixel/140 seconds). This is well below the sky level, and thus not an issue, in all filters except the U filter, where the calibration procedure compensate for this effect.

d) The gains of the electronic channels were stable to abut the 5% level over the long run (except for a few discrete occasions when faulty electronics was replaced). These variations in gain are corrected for in the calibration procedures in the data processing programs, as described in a previous article (Andrews 2003)

7.1. Alignment of the Instrument

During the initial commissioning phase, the plane of the front surfaces of the CCDs had to be aligned to be parallel to the focal plane to within the ± 25 _m depth of field of the telescope. This was done by tilting adjustments around the two relevant rotational degrees of freedom. The method used was to precisely measure the optimal focus position of each individual CCD. Initially, when the CCD plane was not exactly parallel to the telescope focal plane, a systematic tilt in these focal positions was observed. After several iterations of adjusting the tilt angle, the CCD plane was brought into the telescope focal plane well within the depth of field. A check of the adequacy of this alignment is to examine the FWHM of the PSF of the images in the individual CCDs at a single focal position of the entire camera. No systematic variation of the FWHM across the diagonal of the full image plane is observable. This alignment procedure, carried out during commissioning, did not have to be repeated.

The rotational position of each of the four fingers and the readout clocking rate synchronization has to be changed every time the declination of the drift scan is changed. The expected finger position and clocking rates are calculated and set by the control computer for the desired declination. Initially these settings were optimized by varying the finger positions and the clocking rates in small steps around the expected values for different declinations. If the finger position is off, the images are elongated in the north-south direction, and if the clocking rate is off, the images are elongated in the east-west direction. The optimum settings were those that produced the smallest, round images. After some experience with the camera, we learned to trust the computer settings, and this optimization procedure is undertaken only when the images do not have the expected pointlike shape.

7.2. Astrometric Calibration

The astrometric calibration is obtained for each night of data for each CCD by using the USNO catalog. The random error in the RA and Dec measurements of the QUEST data are typically of the order of 0.1 arc seconds, as determined by comparing the centers of objects measured multiple times.

7.3. Seeing Quality

The Full Width Half Maximum (FWHM) of the images for short (~ 10 seconds) point and track exposures are at best 1".6. The image quality is fairly uniform across the full field of view; no significant degradation is noticeable near the edges. A distribution in the image FWHM for a drift scan exposure on a clear photometric night (September 21, 2005) is shown in Figure 20. The peak of this distribution is at 2".1 FWHM. There is a sharp low end cutoff at 1".8. The degradation from the 1".6 seeing for point and track to the 1".8 seeing for drift scanning is consistent with the degradation we expect due to the effects of drift scanning discussed in section 2 above.

7.4. Sky Background

The sky background is a strong function of the filter pass band, increasing from the lowest level in the U filter towards the redder bands. The sky background level is obviously dependent on the phase of the moon. The sky backgrounds during a clear, dark, moonless night are given in Table 5 for the various filters used in the survey.

7.5. Limiting Magnitudes of the Instrument

The limiting magnitude (we use the Vega system) is taken to be the magnitude up to which a single measurement has a signal to noise better than 10 to 1. The limiting magnitudes are different for the different filter pass bands, and depend on the seeing, the clearness of the sky (extinction) and the phase of the moon (sky background). The limiting magnitude can be measured by making a distribution of the one standard deviation error in the magnitude versus the magnitude in a particular filter band on a clear, dark night. A typical distribution of this kind is shown in Figure 21 for the R filter on a dark photometric night. The limiting magnitude is taken to be the magnitude at which the distribution crosses the magnitude error of 0.1 mags line. The limiting magnitudes measured in this way for photometric dark nights are given in Table 5. The limiting magnitudes will obviously be degraded from these values for worse seeing, light cloud covers or brighter phases of the moon. For our 15,000 square degree data sample, using both photometric nights and nights that were not photometric but could be corrected for extinction, the average values of the limiting magnitudes were a few tenths of a magnitude brighter than the numbers given in the table.

### 7.6. Multicolor Photometry

The Drift Scan Survey provides data in seven color bands (see Figure 15) for the entire survey area of 15,000 square degrees. This multicolor survey is very useful in distinguishing different classes of object. A typical application is selecting quasars. Figure 22 shows a U-B versus B-R color-color diagrams for a portion of the sky where there is data from the SDSS, with objects classified by SDSS as stars and quasars with redshifts below 2.2. The quasars are quite clearly separated from the main sequence stars.

## 8. Acknowledgments

We thank the technical staff and the operating crew of the Palomar Observatory for their help in the installation and commissioning of the camera and their assistance in operating and maintaining the instrument over the course of the Palomar QUEST Survey. This work has been supported by the Department of Energy Grant No. DEFG-02-92ER40704, and National Science Foundation Instrumentation Grant, and NSF Operating Grant No. AST-0407297.

Figure 1: The QUEST Large Area Camera at the Prime Focus of the Samuel Oschin Schmidt Telescope at Palomar

Figure 2: Arrangement of the 112 CCD's in the QUEST Large Area Camera

Figure 3: Photograph of the QUEST Camera before installation

Figure 4: Design for the CCD rotating mechanism, showing the pivot points and the computer controlled spiral cams on each finger

Figure 5: Rotation of the CCD support fingers to keep CCD clocking directions tangent to track of stars

Figure 6: Causes of image degradation due to sagitta and clocking synchronization error on a single CCD

Figure 7: Format of the Sarnoff CCDs used in the QUEST Camera

Figure 8: CCD Package

Figure 9: Quantum efficiency of the Sarnoff CCDs. The curve is the design, the points x are actual measurements

Figure 10: Dark Current vs. Temperature for the Sarnoff CCDs. The dark current bottoms out near .02 to .03 e/sec/pixel as indicated by the horizontal lines in the Figure

Figure 11: Read noise of the Sarnoff CCDs

Figure 12: The Camera Dewar: a) face view, b) and c) side views

Figure 13: Liquid Nitrogen Cooling System for the Camera

Figure 14: Camera Shutter and Filter Tray

Figure 15: Characteristics of the filters used on the QUEST Camera. The curves are the measured filter transmissions multiplied by the CCD quantum efficiency

Figure 16: Schematic of the Camera Readout Electronics

Figure 17: The network of computers and programs used to control the Palomar Telescope and the Quest camera. Unshaded boxes show the telescope and camera hardware. Lightly shaded boxes show the hardware controllers. Darkly shaded boxes show the computers. Processes running on each computer are listed inside the respective boxes. Gigabit optical-fiber networks link all of the computers. All connections to the controllers are RS232 or RS485 serial cables. A high-speed optical fiber links Quest7 to the camera electronics

TABLE 1
Properties of the Samuel Oschin Schmidt
Telescope at the Palomar Observatory

______________________________________

| Parameter | Value |
|---|---|
| Clear Aperture Diameter | 1.26 m |
| Focal Length | 3.06 m |
| $f$ ratio | 2.5 |
| Plate scale | 15 _/arc sec |
| Latitude of Observatory | +33°:21 |
| Longitude of Observatory | 116°:51 |
| Elevation | 1706 m |

TABLE 2
Properties of the QUEST Large Area Camera

___________________________________________________

| Property | Value |
|---|---|
| Number of CCDs ……………………….. | 112 |
| For each CCD: | |
|     Pixel Size ……………………. | 13 _ x 13 _ |
|     No of Pixels …………………. | 600 x 2400 |
|     Pixel Size on Sky ……………. | 0.876" x 0.876" |
| Array Size, CCDs ……………………. | 4 x 28 |
| Array Size, Pixels …………………….. | 9600 x 16,800 |
| Array Size, cm ………………………… | 19.3 cm x 25.0 cm |
| Array Size on Sky ……………………... | 3.6° x 4.6° |
| Sensitive Area …………………………. | 9.6 square degrees |
| Total Pixels …………………………… | $161 \times 10^6$ |

___________________________________________________

TABLE 3
Properties of the Sarnoff CCDS
________________________________________________

| Property | Value |
|---|---|
| Array size ………………………. | 600 x 2400 |
| Pixel size ………………………. | 13 μ x 13 μ |
| Physical size ……………………. | 8.9 mm x 34.6 mm |
| Read noise ………………………. | 6 electrons |
| Dark Current | |
|     At room temperature…… | 500 pA/pixel |
|     At -100° C …………….. | 0.1 electrons/sec |
| Quantum Efficiency ……………. | 95% at Peak of 6000 A |
| Antireflection Coating ………….. | 450 nm thick Bismuth Oxide |

TABLE 4
QUEST Readout Electronics Specifications

| Property | Value |
| --- | --- |
| Electronics Read Noise | <1 ADU |
| Read Rate | 100 kpixel/sec |
| Max. Line Rate | 80 hz |
| Typical Complete Mosaic Read Time | 30 sec |
| Maximum Data Throughput | 5.5 Mbit/s |

TABLE 5
Characteristics of the Instrument

___________________________________________

| Filter | Sky Background e/pixel/140 sec | Magnitude Limit at S/N = 10/1 |
|---|---|---|
| U | 20 | 17.5 |
| B | 300 | 21.0 |
| R | 1500 | 21.0 |
| I | 2300 | 20.6 |
| r | 1400 | 21.2 |
| i | 1000 | 20.8 |
| z | 700 | 19.6 |

___________________________________________

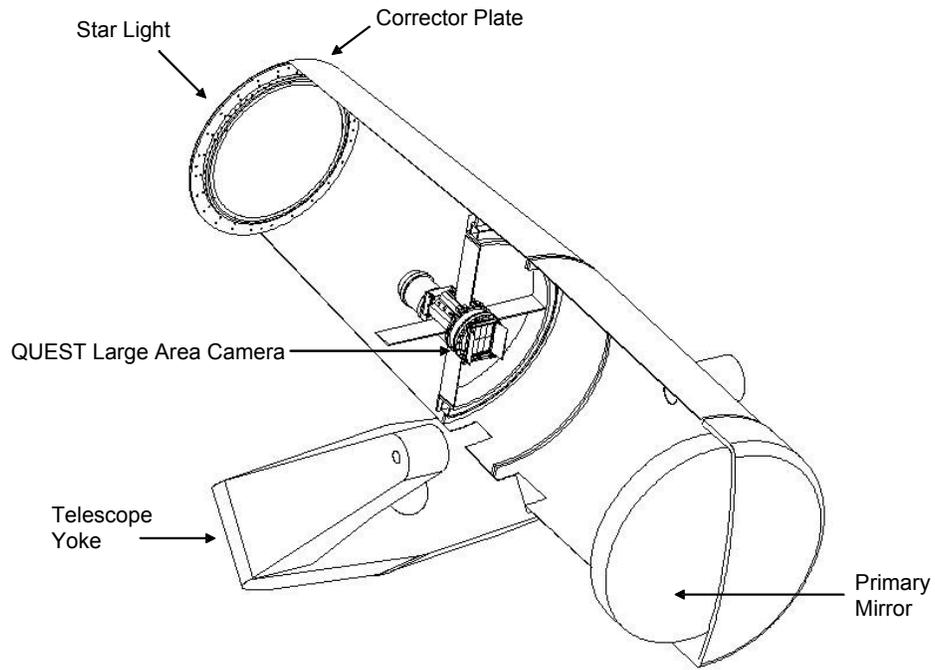

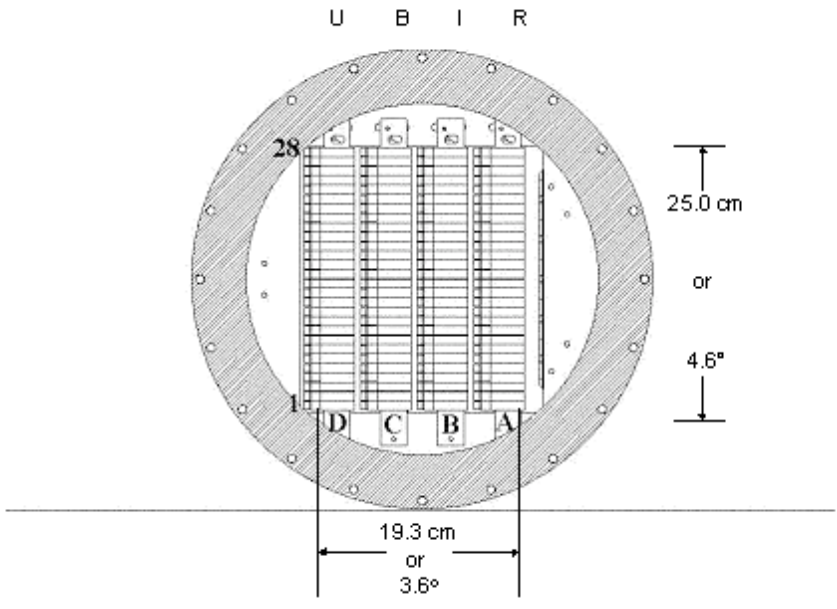

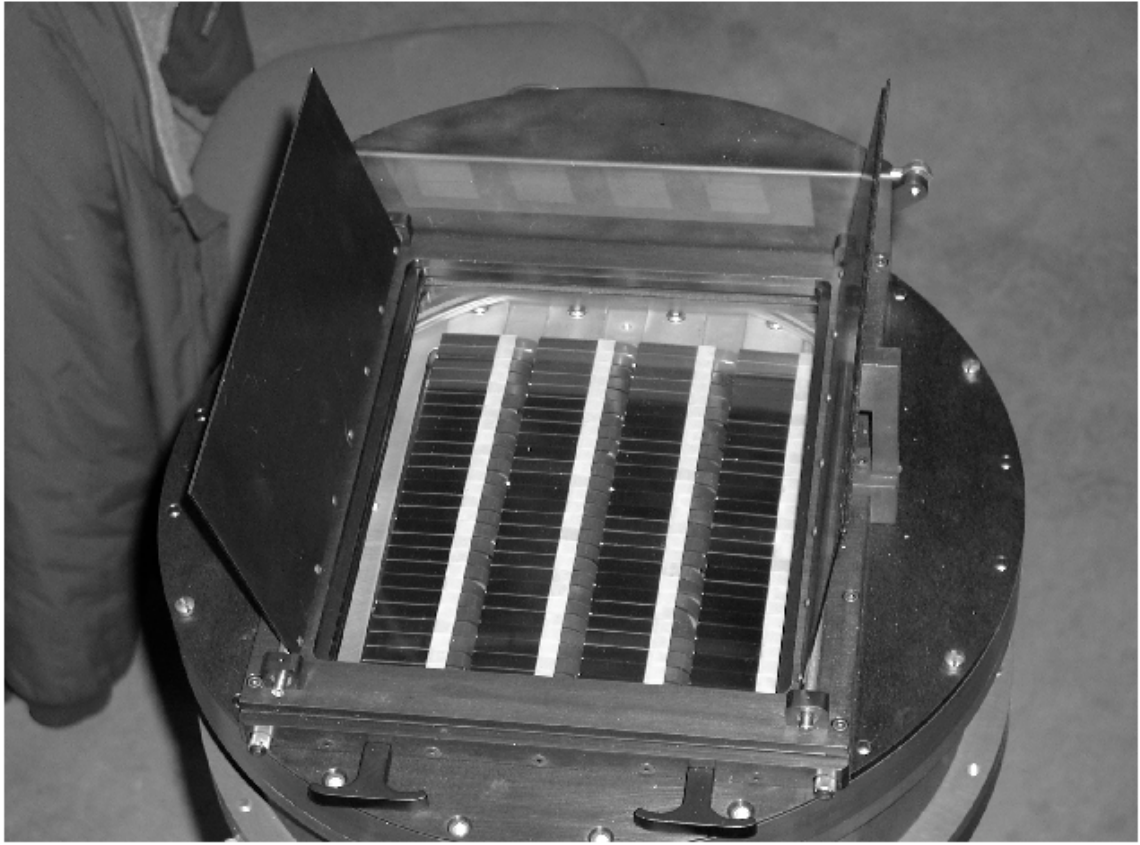

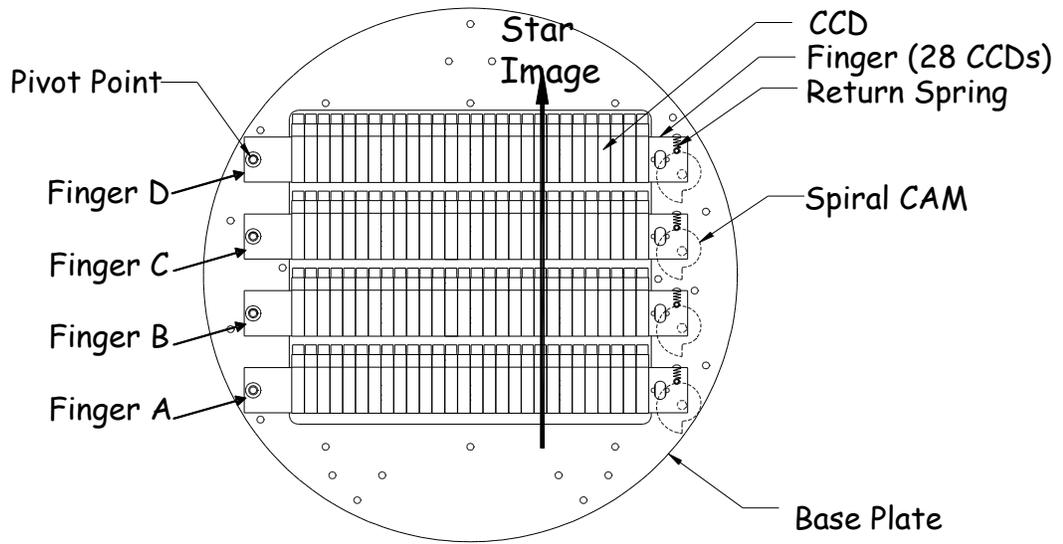

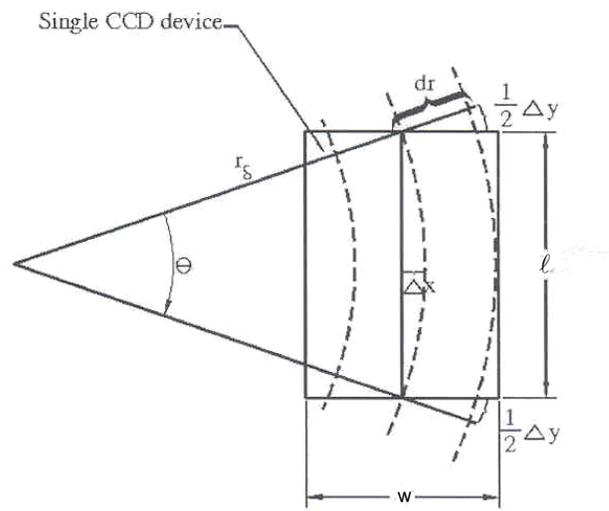

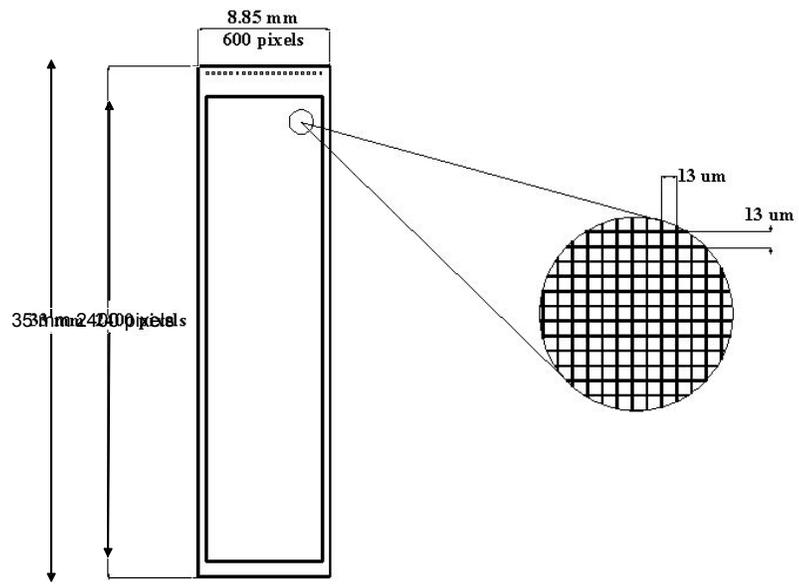

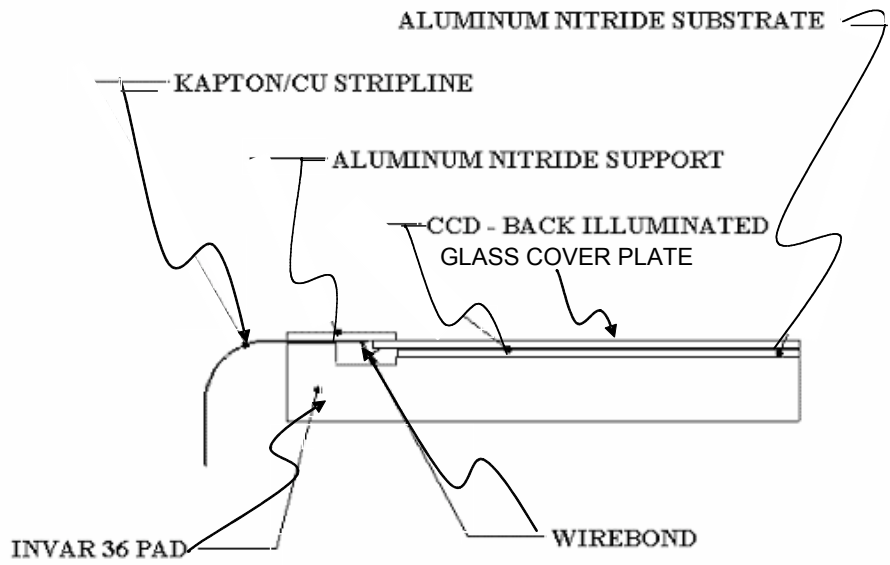

Figure 3.2 CCD Package

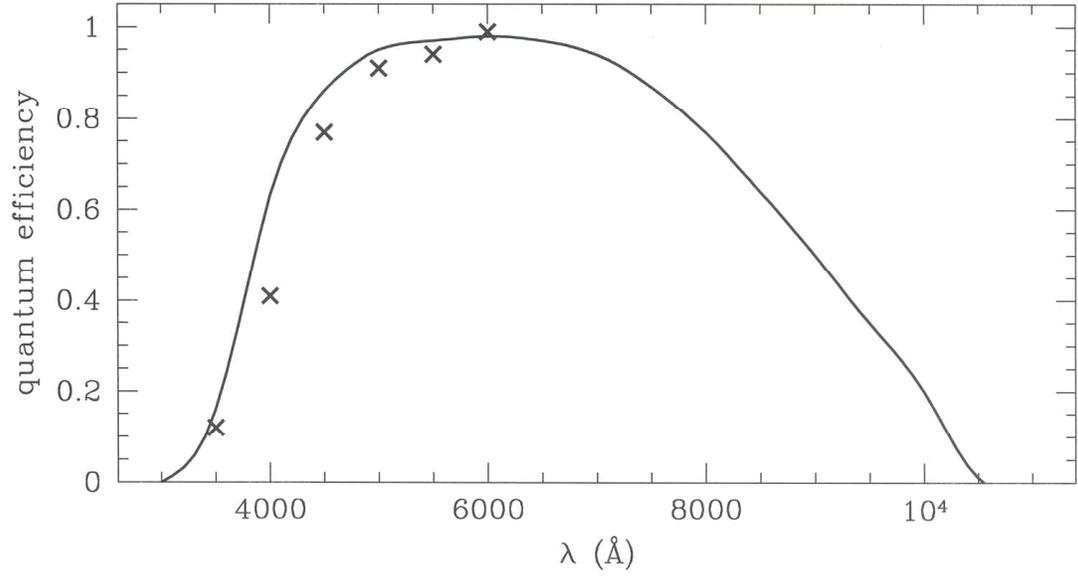

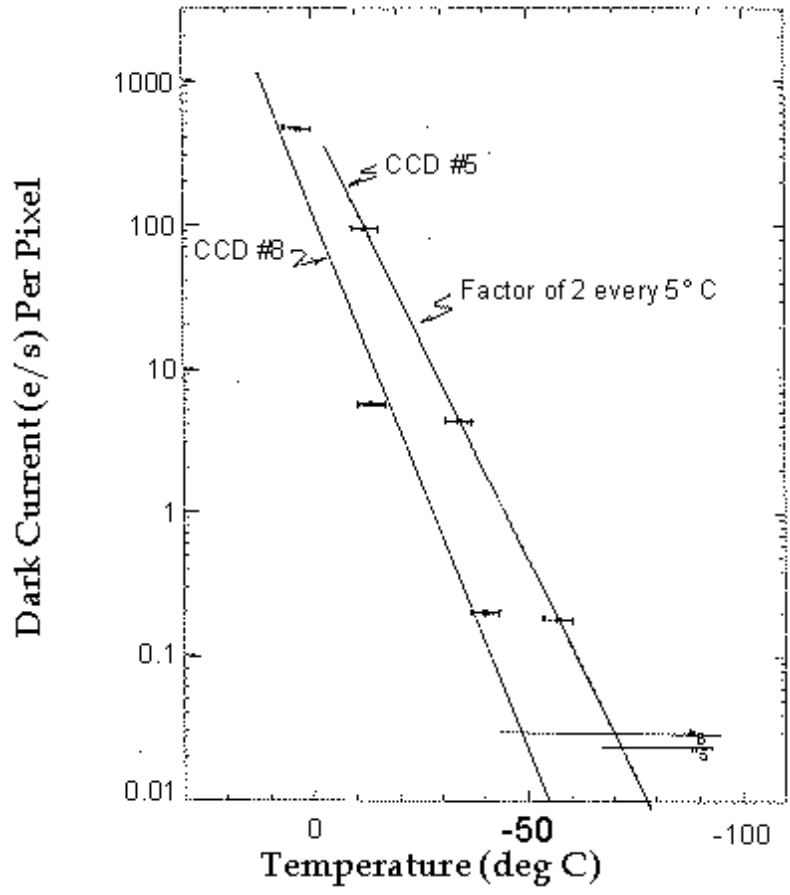

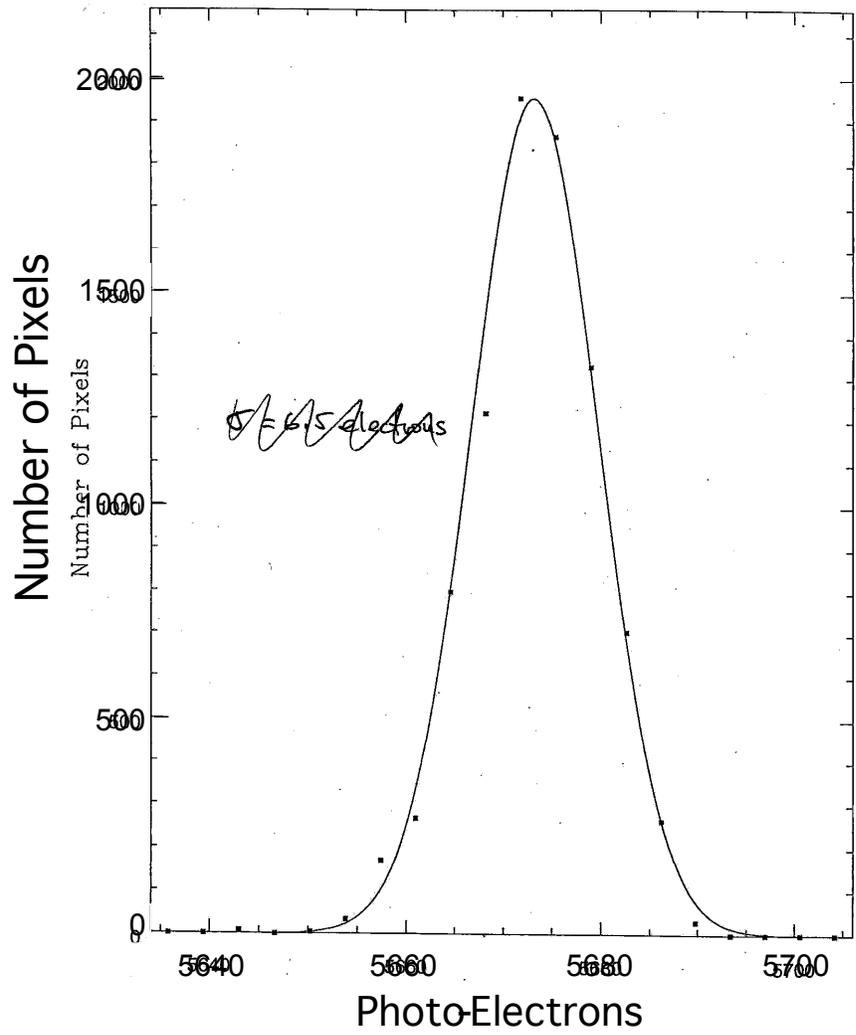

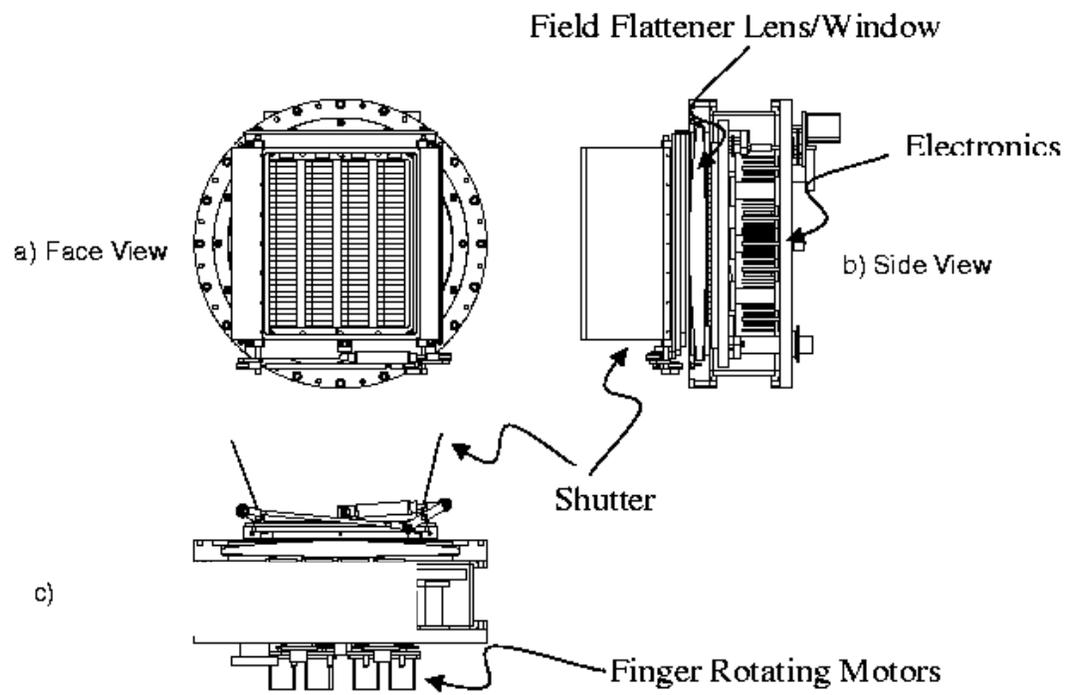

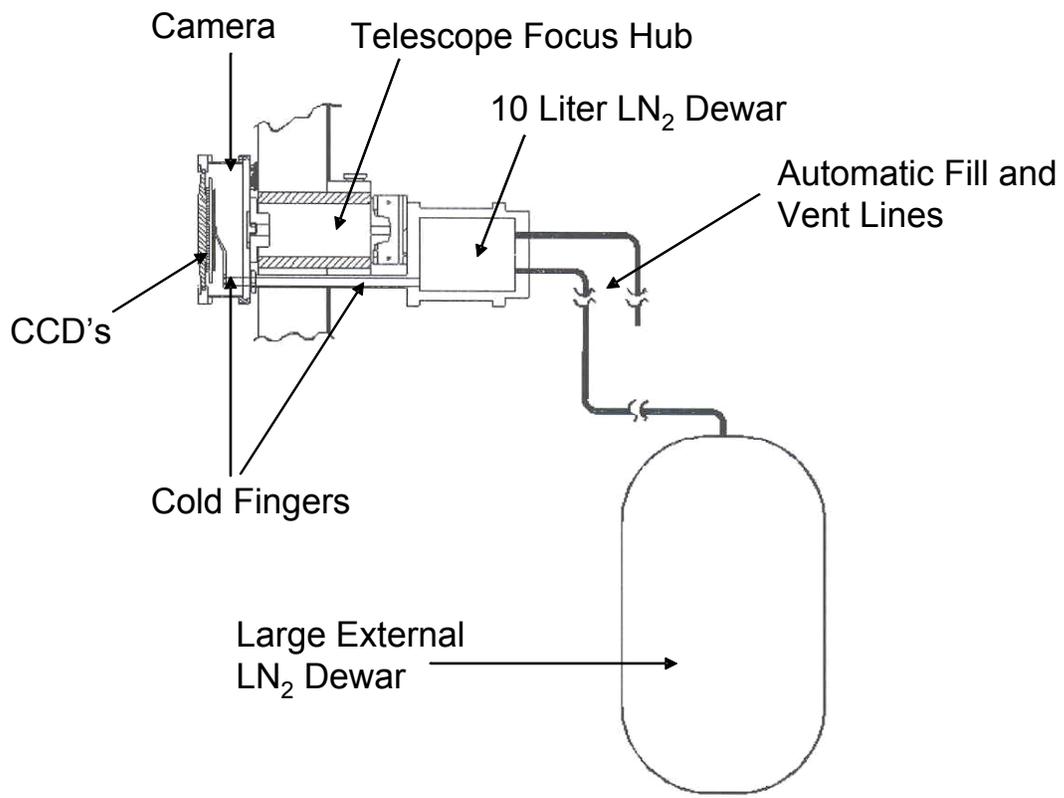

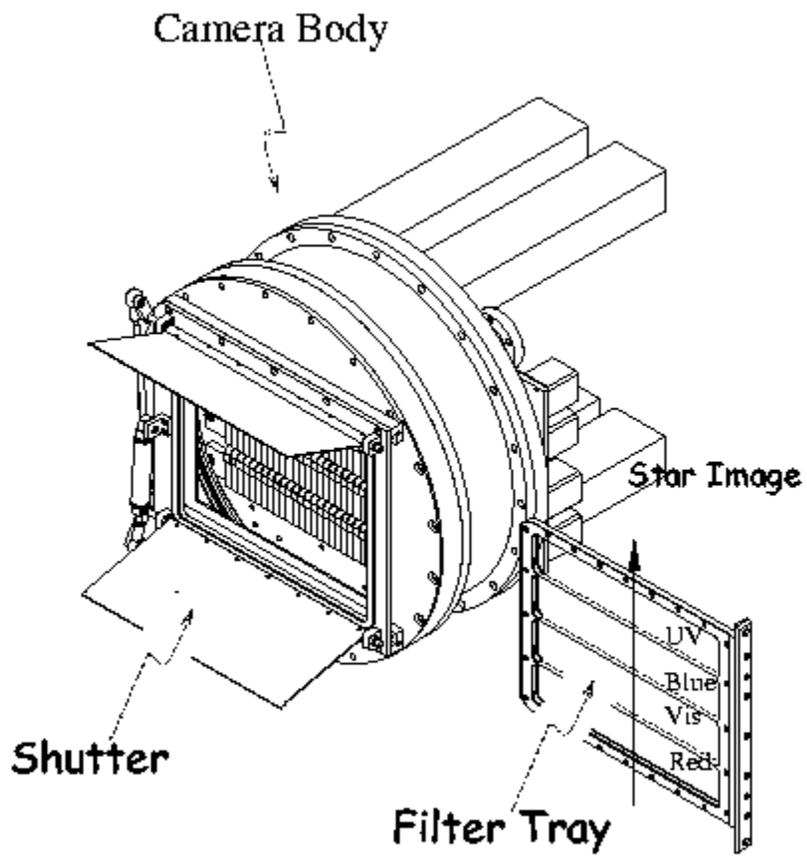

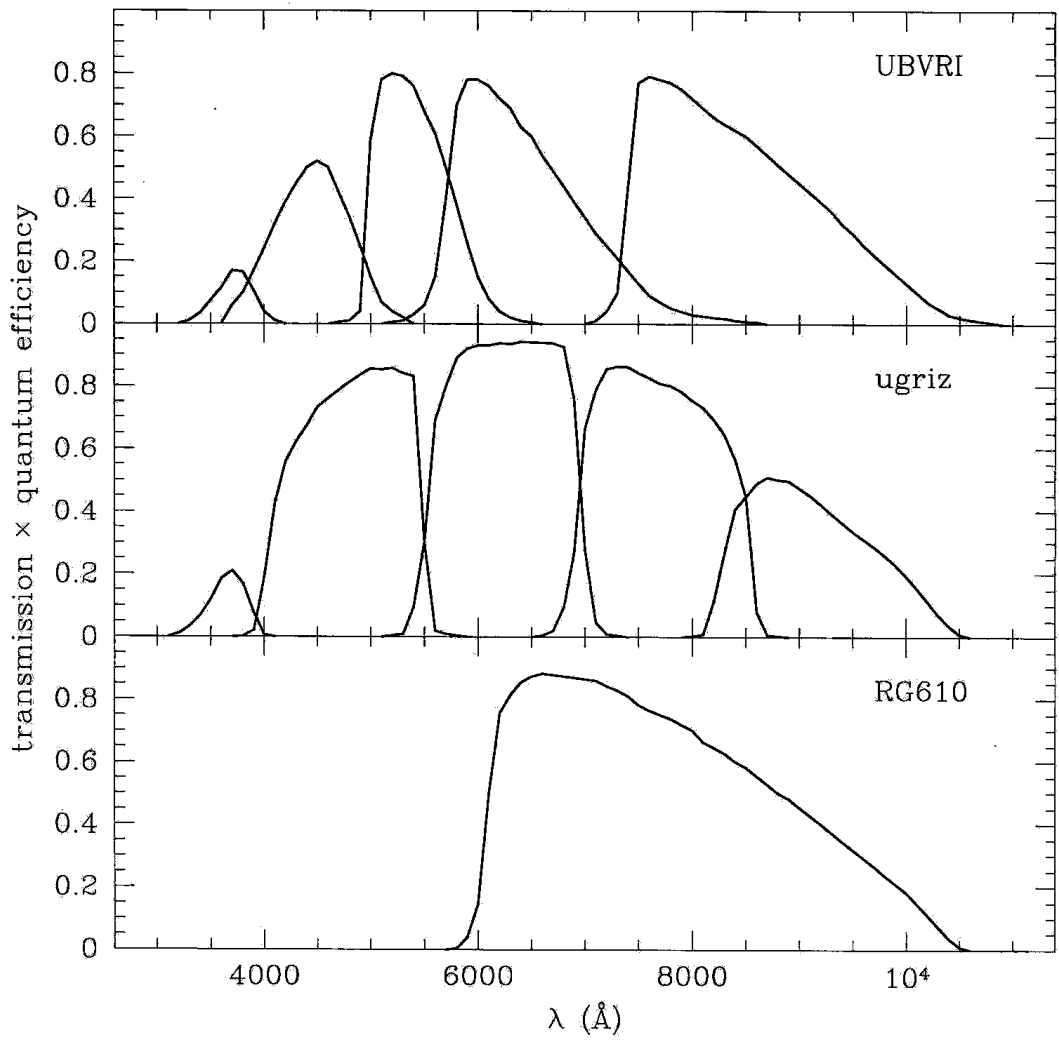

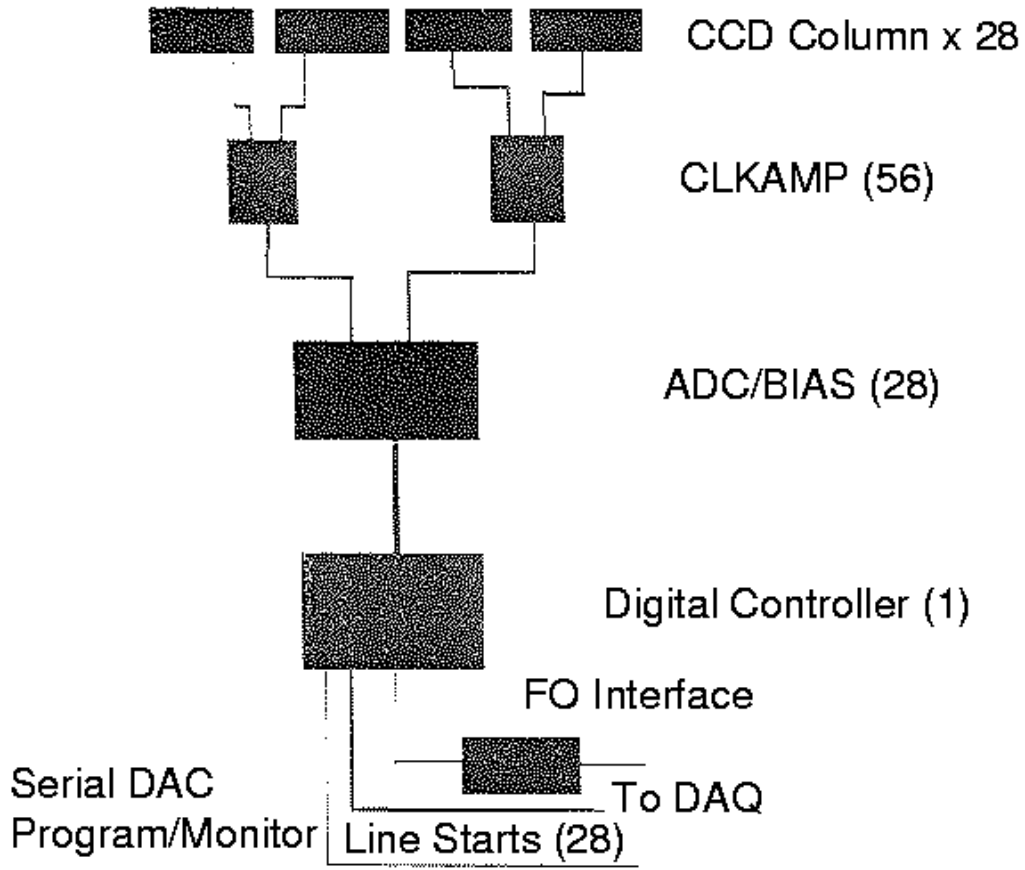

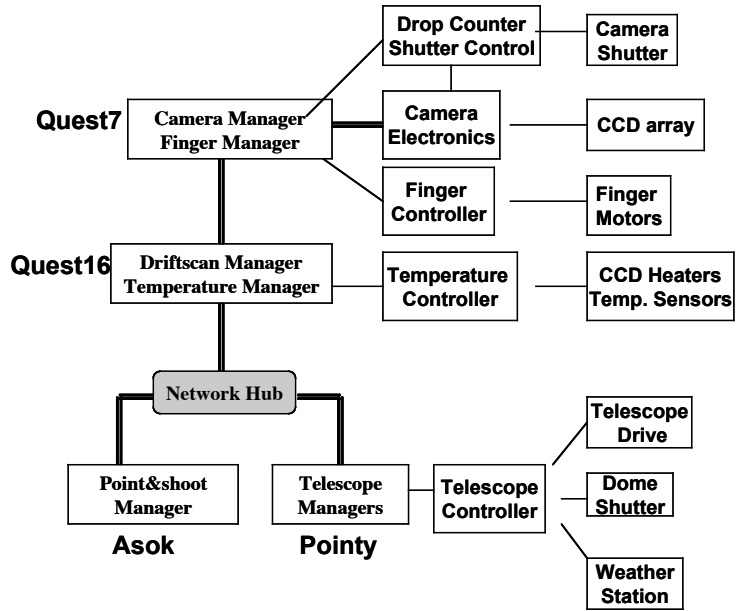

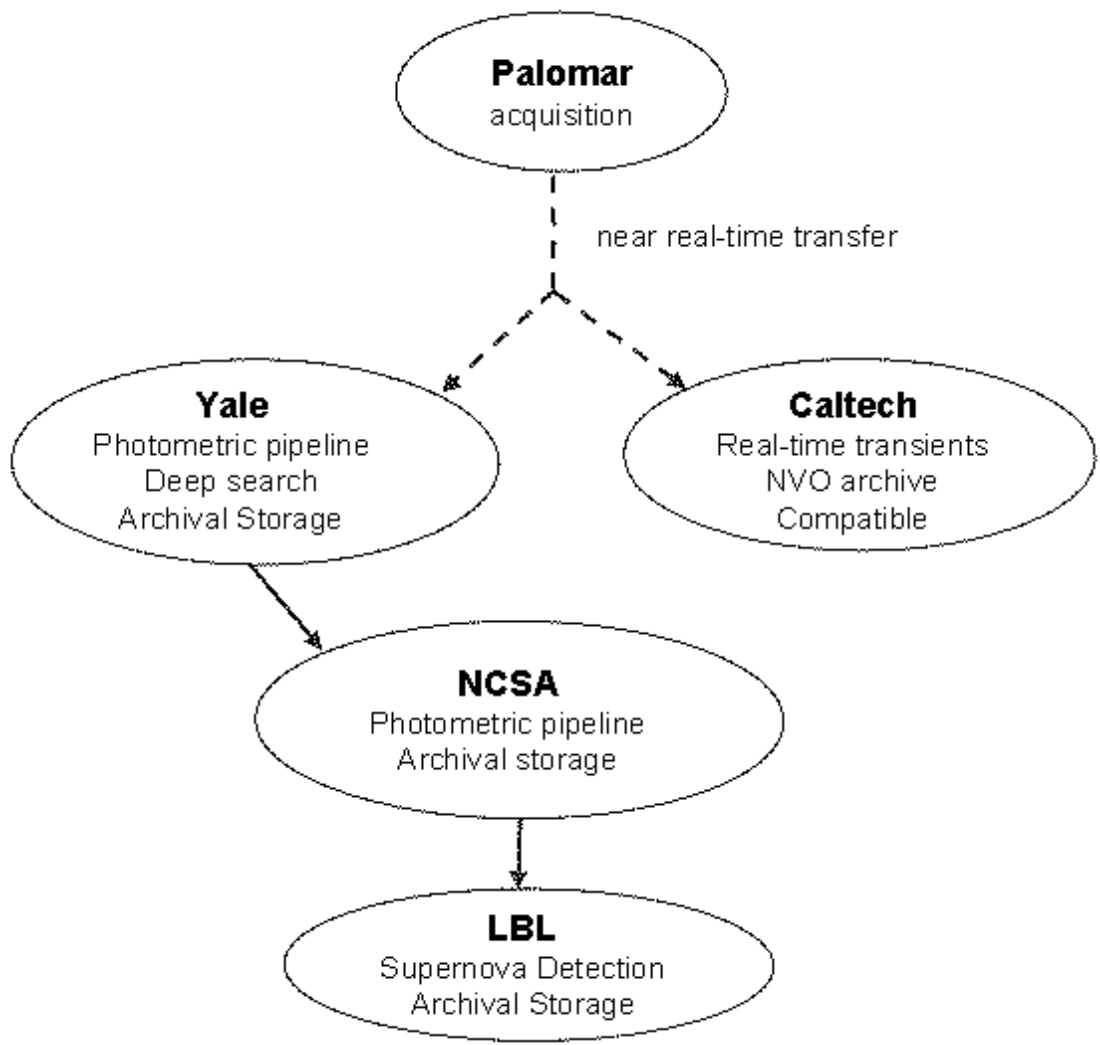

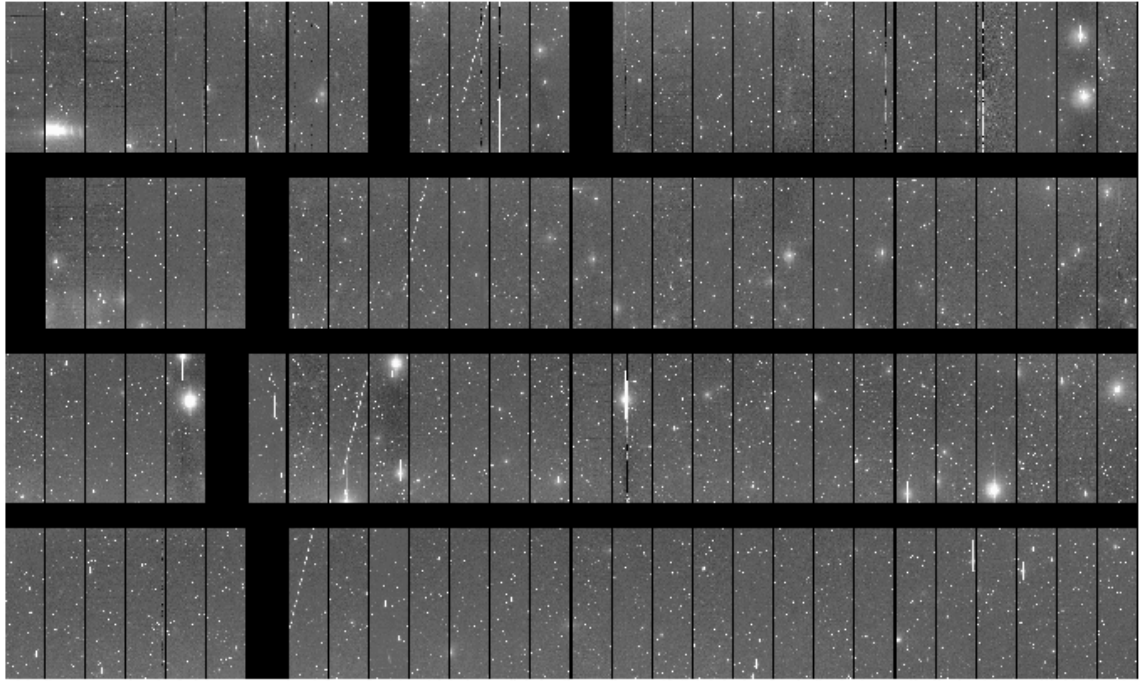

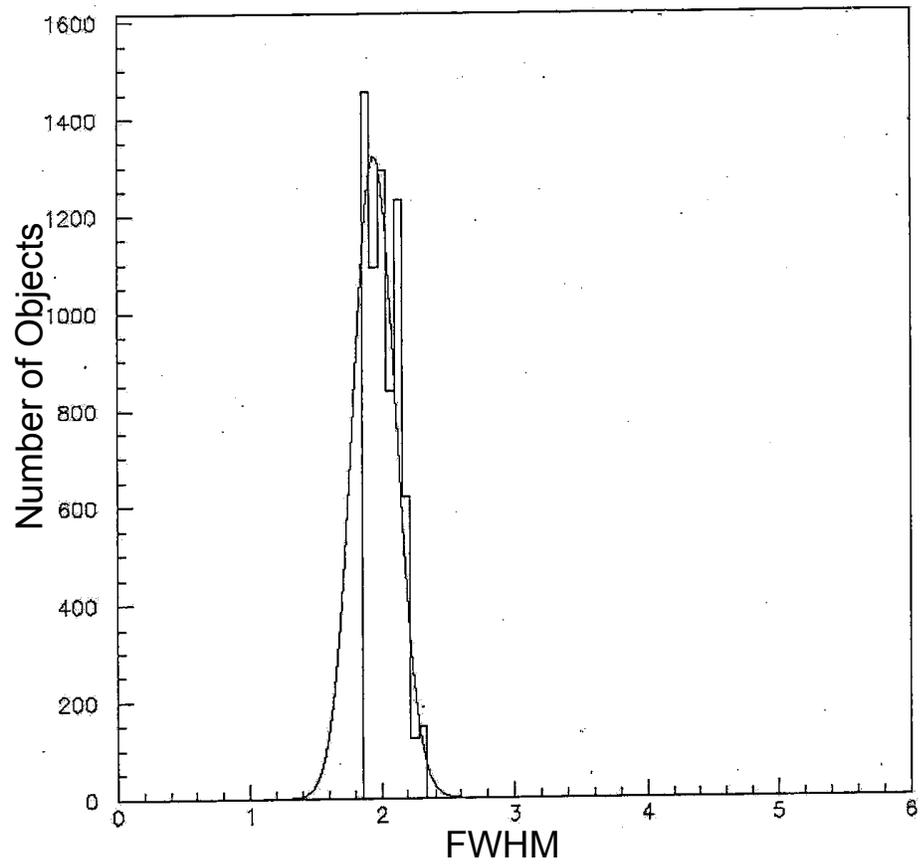

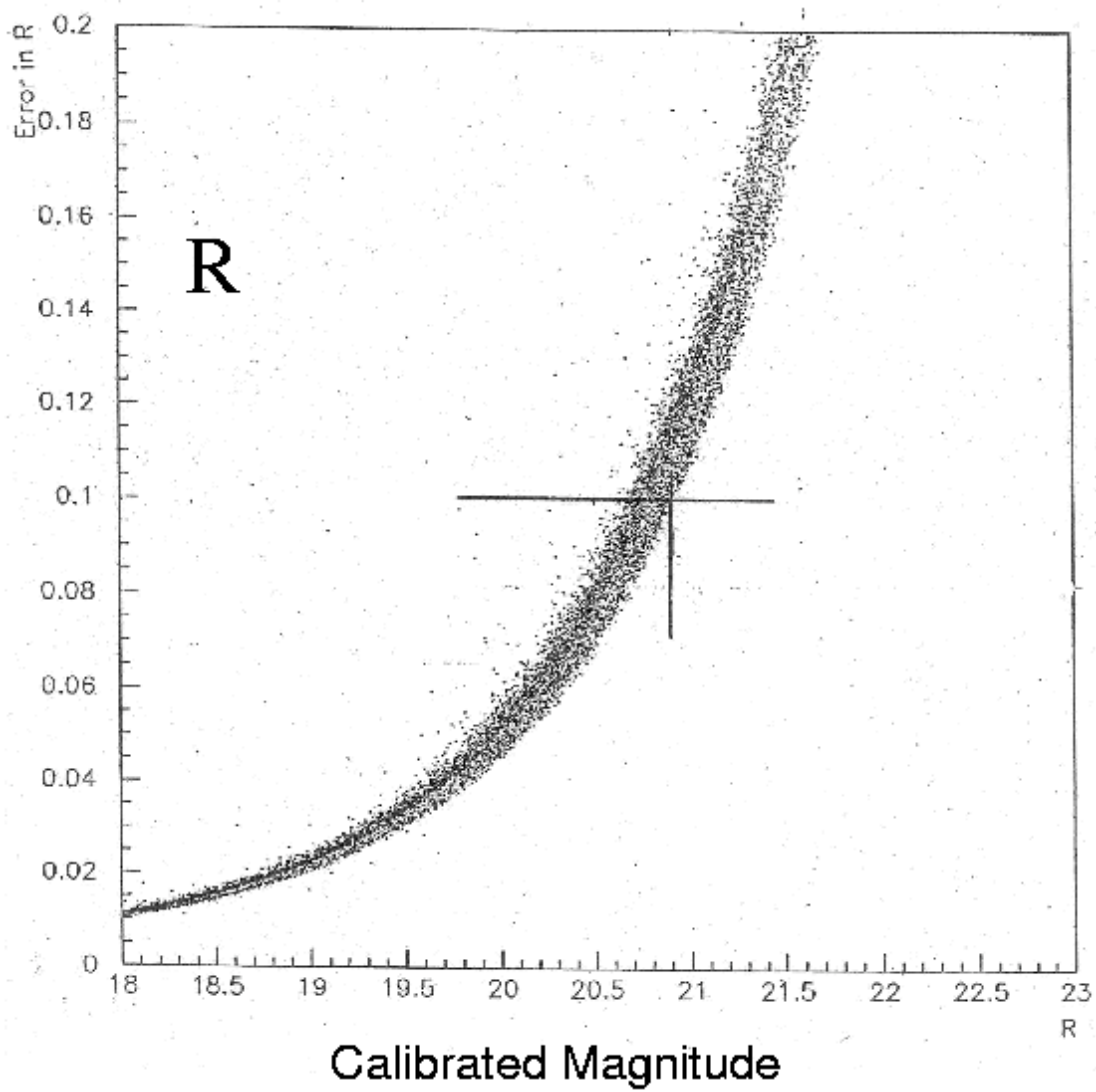

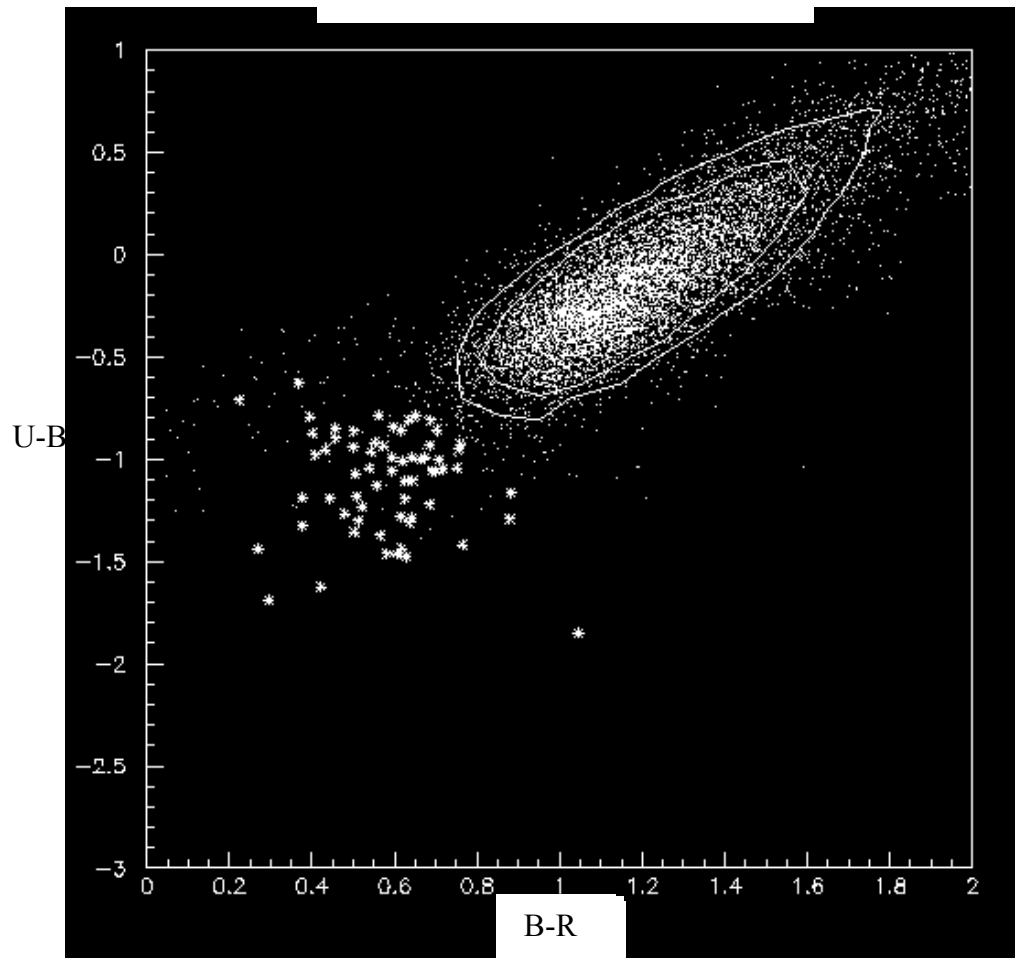